\documentclass[twocolumn,prb]{revtex4} 

\usepackage{amsmath}    
\usepackage{graphicx}  
\begin{document}

\title{Comments on the tethered galaxy problem}

\author{William J. Clavering}
\affiliation{Queen Mary, University of London, School of Mathematical Sciences, London, UK, E1 4NS}
\email{w.j.clavering@qmul.ac.uk}
\date{\today}

\begin{abstract}
In a recent paper Davis et al.  make the counter intuitive assertion that a galaxy held `tethered' at a fixed distance from our own could emit blueshifted light. Moreover, this effect may be derived from the simplest Friedmann-Robertson-Walker (FRW) spacetimes and the $\Omega_{M}=0.3$, $\Omega_{\Lambda}=0.7$ case which is believed to be a good late time model of our own universe.\

In this paper we recover the previous authors' results in a more transparent form. We show how their results rely on a choice of cosmological distance scale and revise the calculations in terms of observable quantities which are coordinate independent. By this method we see that, although such a tethering would reduce the redshift of a receding object, it would not do so sufficiently to cause the proposed blueshift. The effect is also demonstrated to be much smaller than conjectured below the largest intergalactic scales. We also discuss some important issues, raised by this scenario, relating to the interpretation of redshift and distance in relativistic cosmology.
\end{abstract}

\maketitle


\section{Introduction} 

The homogeneous and isotropic expansion of the universe is described by the FRW spacetimes \cite{rin01}. In these models we may construct a comoving frame in which the spacetime of general relativity is treated as expanding whilst matter is, on average, at rest. If we wish to study independently dynamical objects within such models we should like to quantify the effect of imposing such a cosmological background. We follow Davis et al. \cite{dlw03} and consider the extreme case where a galaxy is endowed with sufficient peculiar velocity to escape the universal expansion, or Hubble flow, and remain at a fixed distance from a comoving observer. We study how light from such a galaxy would be redshifted and modify the previous calculations which suggest that a receding source could be significantly blueshifted. We show that such effects, though present, are not as great as was previously suggested and only apply at very large distances for which we would not expect proportionally large peculiar velocities. Throughout this paper we shall permit the term redshifted to include frequency shifts in which light is negatively redshifted (blueshift). In particular, the condition for light to be received with zero redshift implies that it is observed at emission frequency.\

In section \ref{tgp} we review the `tethered galaxy problem' as posed by Davis et al. \cite{dlw03}. By removing the explicit redshift dependence from their calculations we recover their results as a combination of cosmic redshift and the special relativistic Doppler shift. We are thus able to demonstrate why the peculiar velocity required in cancelling the cosmic redshift does not correspond to that proposed to tether a galaxy against the Hubble flow.\

We then discuss two problems with this formulation of the problem. Firstly we note that the peculiar velocities do not correspond to a quantity we might regard as worldline velocity except in the special relativistic limit. In section \ref{grz} we construct a general relativistic condition on the 4-velocity of a luminous particle in order that light is received at the fixed spatial origin without redshift.\

Secondly, in section \ref{dist} we discuss the limitations of the distance scale used in the original formulation. Using this measure we show that, for the Milne model under a coordinate transformation, the problem does not, as we might expect, agree with the analogous system in special relativity. Full calculations for this example appear in the appendix. Motivated by this we propose recasting the problem in terms of a theoretically observable quantity, the radar distance.\

We construct this new scenario in section \ref{radar} and propose a method for solving the system. In section \ref{quant} we compare the phenomenological results with a physical model. We see that the effects of the tethered galaxy problem persist although they are comparatively small below scales of $10^4 - 10^5 Mpc$.
\section{The tethered galaxy problem}\label{tgp} 
We consider FRW spacetimes governed by the metric
\begin{equation}
ds^{2}=-c^2dt^{2}+a(t)^{2}\left[d\chi^{2}+\eta^2(\chi)(d\theta^2+\sin^2\theta d\phi^2)\right], \label{frw1}
\end{equation}
where $\eta=\sin\chi, \chi$ or $\sinh\chi$ when the curvature is positively curved, negatively curved or flat respectively. Field equations for such a universe dominated by matter and the cosmological constant reduce to the Friedmann equation \cite{rin01} for $a(t)$. We express this equation in terms of the normalised mass density $\Omega_M=8\pi G \rho_{M,0}/3 H_{0}^2$ and cosmological constant $\Omega_{\Lambda}=\Lambda/3H_0^2$ so that $k=(\Omega_{\Lambda}+\Omega_{M}-1)$ characterizes the curvature.
\begin{equation}
a'=\frac{da}{dt}=H_0 \left[1+\Omega_M\left(\frac{1}{a}-1\right)+\Omega_{\Lambda}(a^2-1)\right]. \label{freq}
\end{equation}
Following Davis et al.\cite{dlw03}, we consider radial motion in an FRW universe. The metric (\ref{frw1}) thus reduces to
\begin{equation}
ds^{2}=-c^2 dt^{2}+a(t)^{2}d\chi^{2}. \label{frw2}
\end{equation}

These authors define `proper' distance, $D=a\chi$, as metric distance along the surfaces of homogeneity $t=const$. We henceforth refer to this measure as comoving distance since it is defined in terms of the comoving coordinate system. Treating all galaxies as test particles, we invoke symmetry and choose our galaxy to reside at the origin of the comoving radial coordinate $\chi$. Other galaxies then move at a rate, 
\begin{equation}
 D' =  a'\chi + a\chi',
\end{equation}
where $'$ denotes differentiation with respect to cosmic time, $t$.

A comoving galaxy, whose worldline is characterised by $\chi'=0$, can thus be considered to retreat from us with velocity $v_{rec}=a'\chi$. This recession of comoving observers is commonly referred to as the Hubble flow. Davis et al. \cite{dlw03} propose `tethering' the galaxy, with a peculiar velocity $v_{teth}$, in order to maintain constant comoving distance, i.e. 
\begin{equation}
 D' =  v_{rec} + v_{teth}=0. \label{d0}
\end{equation}
$\chi'$ is therefore chosen to satisfy $v_{rec}=-a\chi'$.\

The authors then suggest that one might expect light emitted by this galaxy to arrive without change in its wavelength, as would certainly be the case in Newtonian cosmology. They then proceed to rebut this conjecture by imposing a zero redshift condition on the worldline of the galaxy and showing that this is not, in general, consistent with the fixed comoving distance condition. Furthermore they find nonlinear relations between the peculiar velocity required to cancel cosmic redshift, $v_{pec}$, and that which acts as a `tether' to the galaxy, $v_{teth}$, parametrising each as a function of redshift. We now recover their results by a more direct method and highlight problems with this approach.\
 
In general relativity cosmological redshift is usually viewed \cite{rin01} as an effect of the expansion of space on the wavelength of light. For two comoving observers in an FRW universe the redshift of a light signal, emitted at $t_{em}$, is given by
\begin{equation}
1+z_{rec}=\frac{a(t_{obs})}{a(t_{em})}.\label{cos}
\end{equation}

We wish to endow the galaxy with a peculiar radial velocity sufficient to cancel this redshift. Davis et al. \cite{dlw03} invoke the special relativistic Doppler shift \cite{rin01}, characterised by
\begin{equation}
1+z_{pec}=\left(\frac{c+v_{pec}}{c-v_{pec}}\right)^{\frac{1}{2}},\label{dop}
\end{equation}
and demand that the two effects cancel,
\begin{equation}
\left(1+z_{rec}\right)\left(1+z_{pec}\right)=1. \label{zred}
\end{equation}
Combining (\ref{cos}),(\ref{dop}) and (\ref{zred}) we obtain 
\begin{equation}
v_{pec}=\frac{c(a_{em}^2-a_{obs}^2)}{a_{obs}^2+a_{em}^2}.\label{vpec}
\end{equation}

This is not, in general, the same as $v_{teth}=-a'\chi$ and it is this result which appears implicitly in the paper of Davis et al. \cite{dlw03}. In Appendix \ref{milvteth} we recover the relationships $v_{pec}=f(v_{teth})$ analytically for the cases of the Milne cosmology ($\Omega_{M}=\Omega_{\Lambda}=0$) and the flat matter ($\Omega_{M}=1, \Omega_{\Lambda}=0$) and cosmological constant dominated ($\Omega_{M}=0, \Omega_{\Lambda}=1$) FRW models.\
There are two issues to be raised with the argument above. Firstly there is a technical problem with applying equation (\ref{dop}) to FRW models. $v_{pec}$, as defined above, does not correspond to a quantity we might treat as worldline velocity. We shall see this from the general relativistic argument of the next section. Secondly we note that the comoving distance used here is not the only distance scale one could choose and, in particular it does not correspond to an observable quantity.\

In section \ref{dist} we shall see that `tethering' a galaxy at a fixed value of $D$ does not coincide with what we might expect from special relativity in the case of an empty FRW model. Hence we suggest that this distance measure is misleading when used in the `tethering' problem and we propose a more suitable alternative in section \ref{radar}.
\section{Zero redshift condition}\label{grz} 
Equation (\ref{dop}) is derived for the Minkowski space of special relativity \cite{rin01} in which $v_{pec}$ can be interpreted, in the obvious way, as velocity with respect to the inertial frame of our stationary observer. If we choose to employ this result in a general relativistic model, however, we must accept that $v_{pec}$ will not, except in the Minkowski limit, correspond to the conjectured velocity $v_{teth}=-a\chi'$ in comoving coordinates. This method, therefore, fails to give a meaningful description of zero-redshifted observers in FRW cosmology. We now outline the general relativistic version of this argument from which we can recover the 4-velocity of a zero-redshifted observer and hence the corresponding worldline.\

The frequency of a light signal with wave vector $k^a$ measured by an observer with 4-velocity $u^a$ is given by \cite{wal84}
\begin{equation}
\omega=-k_a u^a.\label{omega}
\end{equation}
Therefore, in order that the signal is received at its emission frequency by a second observer, whose 4-velocity is $v^a$, we require $\omega_{obs}=\omega_{em}$, hence
\begin{equation}
k_a u^a |_{t_{obs}}=k_a v^a |_{t_{em}}.\label{cred}
\end{equation}

We again consider an FRW universe with the observer at the origin of comoving coordinates, $u^a=(\frac{1}{c},0,0,0)$, and a radial light signal, $k^a=\left(\frac{1}{a},\frac{-c}{a^2},0,0\right)$. Equation (\ref{cred}), with the normalization condition $v^a v_a =-1$, may be solved for the radial 4-vector $v^a=\left(\dot{t},\dot{\chi},0,0 \right)$, where dot denotes differentiation with respect to an affine parameter. This gives the following equation for the gradient of the zero redshift worldline in comoving coordinates;
\begin{equation}
\frac{d \chi}{d t}=\frac{v^\chi}{v^t}=\frac{c}{a_{em}}\left(\frac{a_{em}^2-a_{obs}^2}{a_{em}^2+a_{obs}^2}\right). \label{zero}
\end{equation}

For models in which the Friedmann equation (\ref{freq}) may be solved analytically for $a(t)$ we can integrate back along the light ray
\begin{equation}
\chi_{em}=c \int^{t_{obs}}_{t_{em}}\frac{dt}{a(t)}, \label{intx}
\end{equation}
and express $t_{obs}=t_{obs}(t_{em},\chi_{em})$. We may then write (\ref{zero}) as an ordinary differential equation in $(t_{em},\chi_{em})$ for the worldline of the zero redshifted observer.
\section{Distance measures in FRW cosmology}\label{dist} 
When dealing with relativistic systems it is often useful to think in terms of invariant quantities rather than those which depend on some particular observer, coordinate system or frame of reference. Working in terms of observer dependent quantities can give rise to paradoxical conclusions.
The distance $D=a(t)\chi$, as defined in section \ref{tgp}, is a logical choice when working with the comoving coordinates $(t, \chi)$. Indeed, the following argument gives a strong impression that it relates to an observable quantity.\
 
Weinberg \cite{win72} appeals to a chain of comoving galaxies lying close together on the line of sight between ourselves and a distant galaxy in an FRW universe. Synchronising clocks to cosmic time $t$, observers in each galaxy each measure the distance to their neighboring galaxies. The comoving distance is then the limit of the sum of these distances,
\begin{equation}
D(t)=\int_{0}^{\chi_{1}}\sqrt{g_{\chi\chi}}d\chi=\int_{0}^{\chi_{1}}a(t)d\chi=\chi_{1}a(t),
\end{equation}

Davis et al. \cite{dlw03}, following Rindler \cite{rin01} further assert that the comoving astronomers in this chain should each lay rulers end to end to achieve this measurement. However, we should not allow ourselves to be carried away by this image. In Newtonian physics the idea of laying down a ruler between some pair of points in space makes perfect sense, regardless of whether the scales involved make it physically practical. In general relativity, however, we cannot typically define such an idealized rigid body to act as a ruler (see Trautman \cite{tra64}) neither could one hope to obtain the infinite number of observers needed to obtain the limit in Weinberg's gedanken experiment. We suggest here that comoving distance should be regarded as a measure motivated by the coordinate choice rather than an observable quantity.\

Consider now the tethered galaxy problem in the Milne universe. This is the empty case of FRW with zero cosmological constant ($\Omega_{M}=\Omega_{\Lambda}=0$) for which $a(t)= H_0 t/t_0=A_0 t$. The coordinate transformation
\begin{equation}
(T,X)=\left(ct\cosh\left(\frac{A_0 \chi}{c}\right),ct\sinh\left(\frac{A_0 \chi}{c}\right)\right) \label{TX}
\end{equation}
reduces the line element to that of $(1+1)$ Minkowski space, $ds^{2}=-dT^2+dX^2$. In these coordinates geodesics take the form of straight lines and special relativity holds at all scales. We would expect to define a tethered galaxy as moving along one of the timelike geodesics  $X=constant$. Indeed light received from such a galaxy would be seen with zero redshift. However, the family of observers defined by $D'=0$ follow trajectories 
\begin{equation}
A_0 t \chi=(T^2-X^2)^{\frac{1}{2}} \tanh^{-1} \left(\frac{X}{T} \right)=constant.
\end{equation}

Such observers are not seen with zero redshift and the tethering condition does not correspond to the fixing of distance we would expect from special relativity.\

Another choice of distance scale would be to measure metric distance along spatial geodesics. Here, since our aim is to  obtain a relationship with an observable quantity, the redshift of a light signal, we choose a distance measure which is itself the result of an experiment. Radar distance, as discussed in the following section, fulfils this criterion as well as reproducing the expected results for the tethering problem in the Milne case.
 
\section{Tethering with radar distance}\label{radar} 
Radar distance \cite{rin01} is calculated by measuring the proper time between emission at $\tau'_{obs}$ and observation at $\tau_{obs}$ of a light signal reflected from a distant object
\begin{equation}
d_{rad}=\frac{\tau_{obs}-\tau'_{obs}}{2},\label{rad}.
\end{equation}

For a comoving observer $\dot{\chi}=0$ in an FRW universe, as is considered in the tethering problem, cosmic time $t$ is equivalent to proper time $\tau$. We now proceed to investigate the dynamics of a galaxy tethered at fixed radar distance in the comoving coordinates. 

\begin{figure}
 \begin{center}
 \fbox{\includegraphics[width=4cm]{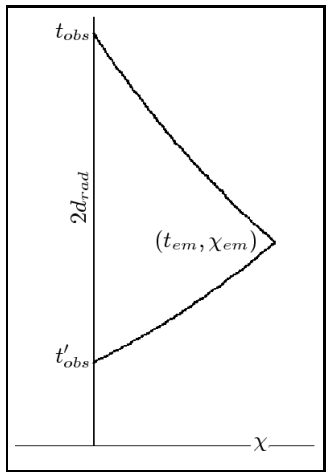}}
 \caption{\footnotesize{Worldline of the light signal required to calculate radar distance to an object}
 \label{radpic}}
  \end{center}
 \end{figure}

Following Jennison and McVittie \cite{jmv74} we consider a light signal emitted from an observer at the spatial origin at time $t'_{obs}$, received by the second observer at $t_{em}$, immediately reflected and returned to the first observer at a time $t_{obs}$. The path of such a light ray, in $(t,\chi)$ coordinates, is illustrated in figure \ref{radpic}.

Integrating along the light path on the outward and return journey gives
\begin{equation}
\chi_{em}=c \int_{t'_{obs}}^{t_{em}}\frac{dt}{a(t)}=c \int_{t_{em}}^{t_{{obs}}}\frac{dt}{a(t)}. \label{radX}
\end{equation}

In order to tether the second observer, we demand $t'_{obs}=t_{obs}-2d_{rad}$, where $d_{rad}$ is the constant radar distance. Equation (\ref{radX}) then provides a simultaneous equation in $t_{em}$ which can be solved for given $t_{obs}$ and $d_{rad}$. We thus recover $\chi_{em}$ as the solution to (\ref{radX}). For models in which the Friedmann equation (\ref{freq}) may be solved analytically, we can obtain a parametric expression for $(\chi,t)$ along the tethered worldline. Otherwise this worldline can be approximated by interpolating the numerical solutions $\left(\chi(t_{obs},d_{rad}),t(t_{obs},d_{rad})\right)$ for fixed $d_{rad}$ and variable $t_{obs}$.
\section{Quantifying the effect over intergalactic scales}\label{quant} 
We now have methods to compute the worldlines of zero redshift observers and those tethered by comoving and radar distance. The issue we must address is whether these worldlines are fundamentally different and how such differences manifest themselves on cosmic scales.\

\begin{figure}
\begin{center}
\includegraphics[width=8cm]{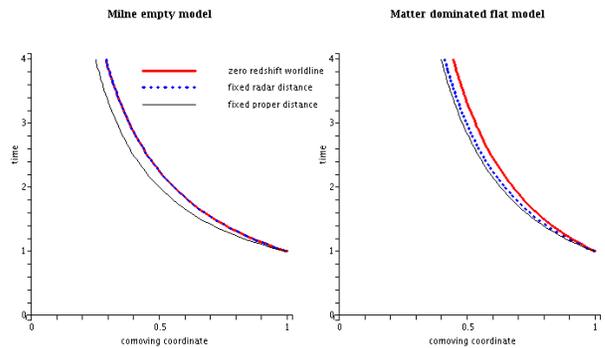}
 \caption{\footnotesize{Worldlines demonstrating the phenomenological differences between the zero redshift condition and the two notions of tethering. In the Milne model the zero redshift and fixed radar distance worldlines coincide.}}
 \label{rrpworl}
 \end{center}
\end{figure}

Our premise, in section \ref{dist}, for preferring a radar distance tether was that it coincides with the zero redshift condition for the Milne model whereas a comoving tether does not. In general, however, all three conditions produce distinct worldlines. A simple example of this is the matter dominated universe, $a(t) \propto t^{\frac{2}{3}}$. We plot this case, along with the Milne model, in figure \ref{rrpworl}. Here we have set the cosmological parameters $c$, $H_0$ and $t_0$ to unity.

Having established that there is a phenomenological difference we now quantify this for the case of galaxies, at typical astronomical distances, tethered by radar distance. We consider a flat universe dominated by the cosmological constant for which $a(t)=e^{H_0(t-t_0)}$ and an observer who receives a signal from a radially tethered galaxy at the current time, $t_{obs}=t_0$. Since $a_{obs}=1$, we may deduce from (\ref{radX}) that 
\begin{equation}
a_{em}=\beta a_{obs}=\beta =\frac{c}{\chi_{em}H_0+c} \label{k}
\end{equation}
in terms of the comoving coordinate at emission, $\chi_{em}$. We then follow the method of section \ref{radar} to obtain the worldline $(t_{em},\chi_{em})$ and hence the velocity
\begin{equation}
\frac{\dot{\chi}}{\dot{t}}=\frac{d\chi}{dt}=\frac{c(\beta -1)}{\beta }.
\end{equation}
Solving the normalization condition $v^a v_a=-1$ for the 4-velocity of the galaxy $v^a=(\dot{t},\dot{\chi},0,0)$ gives
\begin{equation}
\dot{t}=\frac{1}{c}\left(\beta (2-\beta )\right)^{-\frac{1}{2}}.
\end{equation}
Finally we combine this with (\ref{omega}) and express the observed redshift as a function of the comoving separation at emission,
\begin{equation}
\frac{\omega_{em}}{\omega_{obs}}=\frac{c a_{obs}\dot{t}+a_{em}a_{obs}\dot{\chi}}{a_{em}},
\end{equation}
\begin{equation}
z=\frac{\omega_{em}}{\omega_{obs}}-1=\frac{1}{\sqrt{\left(\frac{c}{H_0 \chi_{em}}+c\right)\left(2-\frac{c}{H_0 \chi_{em}}-c\right)}}.
\end{equation}

\begin{figure}
\begin{center}
\includegraphics[width=8cm]{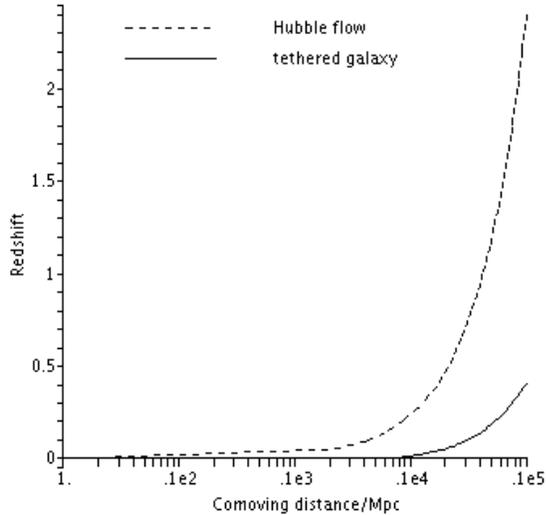}
 \caption{\footnotesize{Redshift of a tethered galaxy as compared to that of a galaxy following the Hubble flow at emission for initial separations in the range $1 Mpc - 10^5 Mpc$. $\chi$ is plotted logarithmically.}}
 \label{ten5}
 \end{center}
 \end{figure}

In figure \ref{ten5} we plot this redshift compared to that of a particle following the Hubble flow at the same emission point.We see that the redshift difference is comparatively small below an initial separation of around $10^4 Mpc$. We also note that the differences between the tethering condition and zero redshift (horizontal axis) become apparent at this scale. This is typical of the general case where we include suitable values for the cosmological parameters. In particular this includes the model $\Omega_{M}=0.3$, $\Omega_{\Lambda}=0.7$, which is seen as the best fit FRW model for our universe. We should note that in figure \ref{ten5} the comoving tether is coincident with the radar distance tether. This is a feature of the $\Omega_{M}=0$, $\Omega_{\Lambda}=1$ model and is not generic in FRW models. For example, in the Milne case, tethered particles starting together at $(t_0,10^4 Mpc)$  would differ in comoving position by around 16\% after $5 Gyr$. 

\section{Summary}
In this paper we have recast the tethered galaxy problem in terms of radar distance and a redshift expression derived from 4-velocity. Consequently, we now obtain a result for the empty FRW model which agrees with special relativity. Furthermore, since all of our results are expressed in terms of observables directly related to the galactic worldlines, we can directly quantify the physical effect. We show that the act of tethering a galaxy against the expansion of an FRW universe results in a greatly reduced redshift when compared to a coincident comoving source. Moreover this effect is only apparent at large astronomical scales and thus is unlikely to influence nearby systems for which we might expect to observe peculiar velocities comparable to those resulting from the Hubble flow. We should note also that these results do not predict blueshifted tethered sources in a physical expanding FRW universe.

\section{Acknowledgments}
I am grateful to Prof. Malcolm MacCallum for invaluable discussions and guidance and to Dr. Susan Scott for suggesting this topic. The work was funded by a PPARC studentship in the Astronomy Unit at QMUL.

\section{Appendix} 
%
%
%
The calculations in \ref{milradteth} and \ref{milz} show that, for the Milne universe, conditions for an observer to move at fixed radar distance from the origin and to emit light with zero redshift are equivalent. This agrees with the argument of section \ref{tgp} where we observed that the Milne universe may be transformed into a subset of Minkowski space in which the equivalence can be seen trivially. In \ref{milvteth} we recover the results of section III in the paper of Davis et al. \cite{dlw03}. This calculation may be compared with Appendix C of that paper. This is where the original assertion that tethered galaxies could appear significantly blueshifted arises. 
\appendix
\section{Fixed radar distance in the Milne universe} \label{milradteth} 
As in section \ref{dist} we consider the empty FRW model in which the scale factor may be written $a(t)= H_0 t/t_{0}=A_0 t$.
A photon is emitted at $t'_{obs}$, reflected by the tethered galaxy at $t_{em}$ and observed at $t_{obs}$. Along null geodesics $ds^2=0$ so we characterize the path of a photon by 
\begin{equation}
\chi=c\int_{t_{obs}'}^{t_{em}}\frac{dt}{a(t)}=c\int_{t_{em}}^{t_{obs}}\frac{dt}{a(t)} \label{milx}.
\end{equation}
If we require the radar distance to remain fixed for all times then
\begin{equation}
t_{obs}-t_{obs}'=2d. \label{miltd}
\end{equation}
Combining (\ref{milx}) and (\ref{miltd}) we obtain the relation
\begin{equation}
t_{em}^2=t_{obs} (t_{obs}-2d), \label{miltt}
\end{equation}
which we substitute into the right hand side of (\ref{milx}),
\begin{equation}
\chi=\frac{c}{A_0 }\left( \ln(t_{obs})-\ln(t_{em}) \right). \label{milex}
\end{equation}

Rearranging (\ref{milex}) to $t_{obs}=e^{\chi A_0 / c}t_{em}$ and combining with (\ref{miltt}) gives an expression $t(\chi)$ for the worldline for the galaxy tethered by constant proper distance, 
\begin{equation}
t_{em}=\frac{2d}{\sinh(\frac{A_0 \chi}{c})}. \label{milrad}
\end{equation}

This is precisely as we would expect from (\ref{TX}) where we hold $X$ constant and set $2d=\frac{X}{c}$.
\section{Zero redshift in the Milne universe}\label{milz} 
Equation (\ref{zero}) tells us that 
\begin{equation}
\frac{d\chi}{dt}=\frac{c}{A_0 t_{em}}\left(\frac{(A_0 t_{em})^2-(A_0 t_{obs})^2}{(A_0 t_{em})^2+(A_0 t_{obs})^2}\right).
\end{equation}
(\ref{intx}) then gives
\begin{equation}
\chi=c\int_{t_{em}}^{t_{obs}} \frac{dt}{a(t)}=\frac{c}{A_0}\left(\ln(t_{obs})-\ln(t_{em})\right),
\end{equation}
and hence
\begin{equation}
t_{obs}=t_{em}  e^{\frac{A_0}{c} \chi}.
\end{equation}
Combining these two formulas gives us
\begin{equation}
\frac{d\chi}{dt}=\frac{c}{A_0 t_{em}}\left(\frac{1-e^{\frac{2A_0 \chi}{c}}}{1+e^{\frac{2A_0 \chi}{c}}}\right)=\frac{c\tanh\left(\frac{A_0 \chi}{c}\right)}{A_0 t_{em}},
\end{equation}
and this agrees with what we get by differentiating (\ref{milrad}).
\section{Analytic forms of $v_{teth}=f(v_{pec})$ for simple FRW models} \label{milvteth} 
From section \ref{tgp} we consider $v_{teth}=-a'\chi=A_0 \chi$. We again use equation (\ref{milex}) and see that
\begin{equation}
t_{em}=t_{obs}e^{\frac{v_{teth}}{c}}.\label{milteth}
\end{equation}
Substituting into equation \ref{vpec} then gives the relation
\begin{equation}
v_{pec}=c\frac{1-e^{\frac{v_{teth}}{c}}}{1+e^{\frac{v_{teth}}{c}}}=-c \tanh\left(\frac{v_{teth}}{c} \right).
\end{equation}

The results for the flat matter dominated and cosmological constant dominated universes may be recovered by the same technique;
\begin{equation}
v_{pec}= \frac{1-(1-\frac{v_{teth}}{c})^2}{1-(1+\frac{v_{teth}}{c})^2}.
\end{equation}
and
\begin{equation}
v_{pec}= -\frac{16-(2+\frac{v_{teth}}{c})^4}{16+(2+\frac{v_{teth}}{c})^4}.
\end{equation}
%


\end{document}